# ACCOMPLISH THE APPLICATION AREA IN CLOUD COMPUTING

Nidhi Bansal[1]
authornidhibansal@gmail.com

B.Tech (CSE) from Shobhit Institute of Engineering & Technology (SIET) Meerut[1]

Amit Awasthi[2]
awasthi.amit1989@gmail.com

B.Tech (CSE) from Meerut International Institute of Technology (MIIT) Meerut[2]

## Abstract

In the cloud computing application area of accomplish, we find the fact that cloud computing covers a lot of areas are its main asset. At a top level, it is an approach to IT where many users, some even from different companies get access to shared IT resources such as servers, routers and various file extensions, instead of each having their own dedicated servers. This offers many advantages like lower costs and higher efficiency. Unfortunately there have been some high profile incidents where some of the largest cloud providers have had outages and even lost data, and this underscores that it is important to have backup, security and disaster recovery capabilities.
<u>In education field</u>, it gives better choice and flexibility to IT departments than others. The platform and applications you use can be on-premises, off-premises, or a combination of both, depending on your academic organization's needs.
With cloud computing in education, you get powerful software and massive computing resources where and when you need them. Use cloud services to best combine:
*On-demand computing and storage.
*A familiar development experience with on-demand scalability.
*Online services for anywhere, anytime access to powerful web-based tools.

## Keyword

Cloud computing services, cloud applications, Information Technology, education field, replication concept.

## Introduction

Cloud application services or "Software as a Service " deliver software as a service over the Internet, eliminating the need to install and run the application on the customer's own computers and simplifying maintenance and support.

A cloud application is software provided as a service. It consists of the following: a package of interrelated tasks, the definition of these tasks, and the configuration files, which contain dynamic information about tasks at run-time. Cloud tasks provide compute, storage, communication and management capabilities. Tasks can be cloned into multiple virtual machines, and are accessible through application programmable interfaces (API). Cloud applications are a kind of utility computing that can scale out and in to match the workload demand. Cloud applications have a pricing model that is based on different compute and storage usage, and tenancy metrics.

What makes a cloud application different from other applications is its elasticity. Cloud applications have the ability to scale out and in. This can be achieved by cloning tasks in to multiple virtual machines at run-time to meet the changing work demand. Configuration Data is where dynamic aspects of cloud application are determined at run-time. There is no need to stop the running application or redeploy it in order to modify or change the information in this file

A cloud application is a SOA application that runs under a specific environment, which is the cloud computing environment (platform).

SOA is a business model that addresses the business process management, cloud architecture addresses many technical details that are environment specific, which makes it more a technical model.

**Support in education institution:**

- **Cost.** You choose a subscription or, in some cases, a pay-as-you-go plan—whichever works best with your organization's business model.
- **Flexibility.** Scale your infrastructure to maximize investments. Cloud computing allows you to dynamically scale as demands fluctuate.
- **Accessibility.** Help make data and services publicly available without jeopardizing sensitive information.

Improve some of the common challenges.

- **Increased Storage.** Organizations can store more data than on private computer systems.
- **Highly Automated.** No longer do IT personnel need to worry about keeping software up to date.
- **More Mobility.** Employees can access information wherever they are, rather than having to remain at their desks.
- **Allows IT to Shift Focus.** No longer having to worry about constant server updates and other computing issues, government organizations will be free to concentrate on innovation.

**Cloud application as services:**

1. Software as a Service (SaaS) applications provide the function of software that would normally have been installed and run on the user's desktop.
2. Platform as a Service (PaaS) cloud computing provides a place for developers to develop and publish new web applications stored on the servers of the PaaS provider.
3. Infrastructure as a Service (IaaS) seeks to obviate the need for customers to have their own data centers. IaaS providers sell customers access to web storage space, servers, and Internet connections.

**Cloud host**ing is fast becoming the web hosting solution of choice, especially with e-commerce merchants. Instead of being limited by the space and utilities of a physical web server, those who use the cloud to host their shops and take payments and orders through it find that they can scale their server space to meet their needs without paying for unnecessary utilities or running out of space and having to make the hard decision of which items and services need to be cut from their enterprise.

A **micro theory:**

After some time, when we send any query to the system then it will be pass from our personal computer to clouds (Servers).

Figure.1 shows this theory:

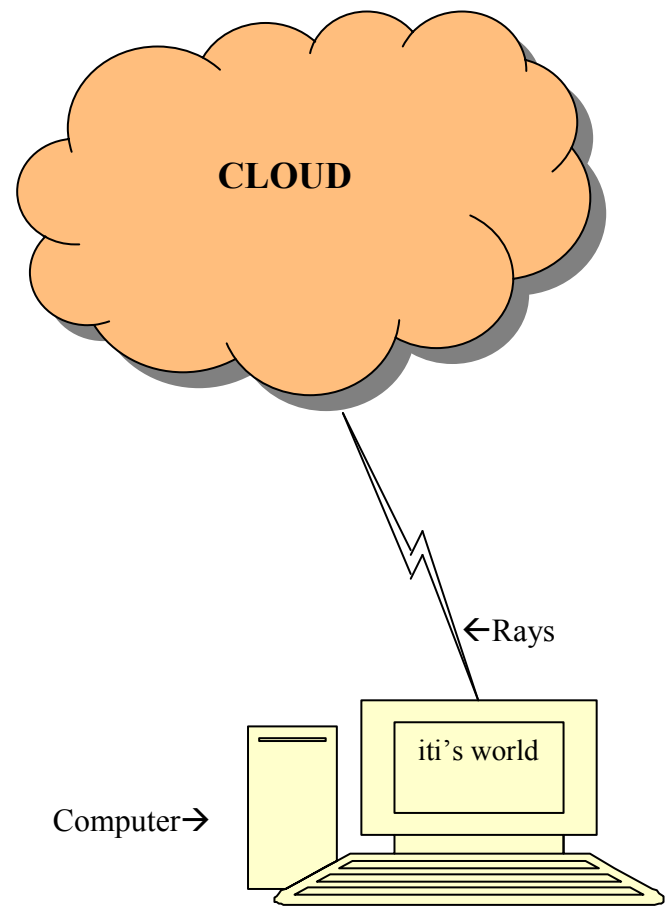

Figure.1

Query 'X' → Personal Computer 'PC' → Process in clouds by Cloud Administrator → Execution Complete, Return to Personal Computer → Result 'Y'

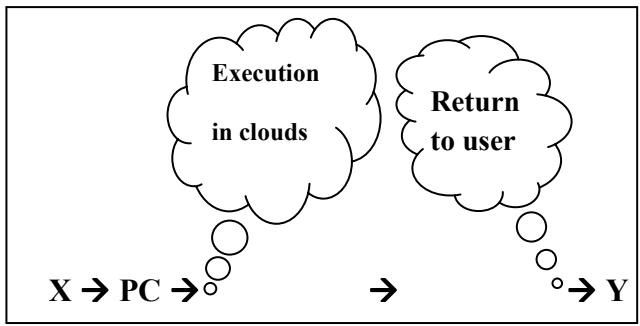

**Functional theory:**
To understand the way to execute query:

**First Step:**
Query will search a port in the whole earth surface to get destination point through these formulas (Figure.2).

**Second Step:**
Query got the server and it will go in execution state.

**Third Step:**
Solution got from user and it will be sent acknowledgement status to the server.

**Fourth Step:**
Update state will be saved by the server.

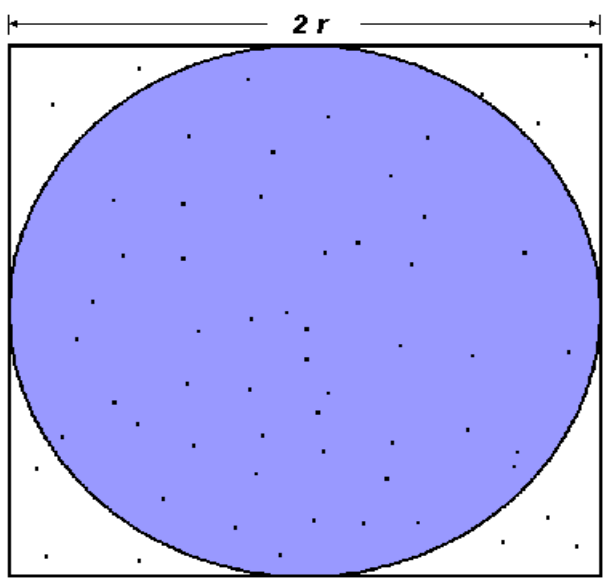

$$A_S = (2r)^2 = 4r^2$$
$$A_C = \pi r^2$$
$$\pi = 4 \times \frac{A_C}{A_S}$$

Figure.2

**Some Technical Security Benefits of the Cloud:**

1. Centralized Data

   * Reduced Data Leakage: this is the benefit I hear most from Cloud providers - and in my view they are right.
   -How many laptops do we need to lose before we get this?
   -How many backup tapes?
   The data "landmines" of today could be greatly reduced by the Cloud as thin client technology becomes prevalent. Small, temporary caches on handheld devices or Notebook computers pose less risk than transporting data buckets in the form of laptops.
   * Monitoring benefits: central storage is easier to control and monitor. The flipside is the nightmare scenario of comprehensive data theft.

2. Incident Response / Forensics

   * Forensic readiness: with Infrastructure as Service providers, we can build a dedicated forensic server in the same Cloud as our company and place it offline, ready for use when needed. We would only need pay for storage until an incident happens and we need to bring it online. We don't need to call someone to bring it online or install some kind of remote boot software - we just click a button in the Cloud Providers web interface. If we have multiple incident responders, we can give them a copy of the VM so we can distribute the forensic workload based on the job at hand or as new sources of evidence arise and need analysis. To fully realize this benefit, commercial forensic software vendors would need to move away from archaic, physical dongle based licensing schemes to a network licensing model.
   * Decrease evidence acquisition time: if a server in the Cloud gets compromised (i.e. broken into), we can now clone that server at the click of a mouse and make the cloned disks instantly available to our Cloud Forensics server. We didn't need to "find"

storage or have it "ready, waiting and unused" - it's just there.

   * Decrease evidence transfer time: In the same Cloud, bit fit bit copies are super fast - made faster by that replicated, distributed file system my Cloud provider engineered for me. From a network traffic perspective, it may even be free to make the copy in the same Cloud. Without the Cloud, we would have to a lot of time consuming and expensive provisioning of physical devices. we only pay for the storage as long as I need the evidence.

   * Eliminate forensic image verification time: Some Cloud Storage implementations expose a cryptographic checksum or hash. For example, Amazon S3 generates an MD5 hash auto magically when you store an object.

* Decrease time to access protected documents: Immense CPU power opens some doors.

-Did the suspect password protect a document that is relevant to the investigation?
You can now test a wider range of candidate passwords in less time to speed investigations.

3. Password assurance testing (aka cracking)

   * Decrease password cracking time: if your organization regularly tests password strength by running password crackers you can use Cloud Compute to decrease crack time and you only pay for what you use. Ironically, your cracking costs go up as people choose better passwords.

   * Keep cracking activities to dedicated machines: if today you use a distributed password cracker to spread the load across non-production machines, you can now put those agents in dedicated Compute instances - and thus stop mixing sensitive credentials with other workloads.

4. Logging

   * "Unlimited", pay per drink storage: logging is often an afterthought, consequently insufficient disk space is allocated and logging is either non-existent or minimal. Cloud Storage changes all this - no more 'guessing' how much storage you need for standard logs.

   * Improve log indexing and search: with your logs in the Cloud you can leverage Cloud Compute to index those logs in real-time and get the benefit of instant search results.
-What is different here?
The Compute instances can be plumbed in and scale as needed based on the logging load - meaning a true real-time view.

   * Getting compliant with Extended logging: most modern operating systems offer extended logging in the form of a C2 audit trail. This is rarely enabled for fear of performance degradation and log size. Now you can 'opt-in' easily - if you are willing to pay for the enhanced logging, you can do so. Granular logging makes compliance and investigations easier.

5. Improve the state of security software (performance)

   * Drive vendors to create more efficient security software: Billable CPU cycles get noticed. More attention will be paid to inefficient processes; e.g. poorly tuned security agents. Process accounting will make a comeback as customers target 'expensive' processes. Security vendors that understand how to squeeze the most performance from their software will win.

6. Secure builds

   * Pre-hardened, change control builds: this is primarily a benefit of virtualization based Cloud Computing. Now you get a chance to start 'secure' (by your own definition) - you create your Gold Image VM and clone away. There are ways to do this today with bare-metal OS installs but frequently these require additional 3rd party tools, are time consuming to clone or add yet another agent to each endpoint.

   * Reduce exposure through patching offline: Gold images can be kept up securely kept up to date. Offline VMs can be

conveniently patched "off" the network.

* Easier to test impact of security changes: this is a big one. Spin up a copy of your production environment, implement a security change and test the impact at low cost, with minimal startup time. This is a big deal and removes a major barrier to 'doing' security in production environments.

7. Security Testing

* Reduce cost of testing security: a SaaS provider only passes on a portion of their security testing costs. By sharing the same application as a service, you don't foot the expensive security code review and/or penetration test. Even with Platform as a Service where your developers get to write code, there are potential cost economies of scale (particularly around use of code scanning tools that sweep source code for security weaknesses).

## Conclusion

We **swear** that, these are only some of the uses that cloud technology currently supports. When it comes to the future, it is obvious that this will become the way of life for those who rely on vast amounts of data and require portability across systems and devices. Cloud services offer a less expensive and far more versatile experience for users to work with their information and to provide services for others on short notice. Also, it decentralizes data storage which will assist the users in feeling far more secure about their data and give one more tool against corrupted hard drives and other accidents of nature. As time goes on, it is certain that the cloud will become integrated with just about every type of activity that takes place on the Internet.

**Result: Process to find data in clouds will be complex but it's very helpful.**

## Acknowledgements

The researchers wish to express their deepest gratitude and warmest appreciation to the following people, who, in any way have contributed and inspired the researchers to the overall success of the undertaking:

•To our parents who have always been very understanding and supportive both financially and emotionally

•and above all, to the Almighty God, who never cease in loving us and for the continued guidance and protection.

## References

1. "An example of a 'Cloud Platform' for building applications". Eccentex.com. Retrieved 2010-08-22.
2. "The Internet Cloud". Thestandard.com. Retrieved 2010-08-22.
3. "Economies of Cloud Scale Infrastructure". Cloud Slam 2011. Retrieved 13 May 2011.
4. "Distributed Application Architecture". Sun Microsystem. Retrieved 2009-06-16.
5. Danielson, Krissi (2008-03-26). "Distinguishing Cloud Computing from Utility Computing". Ebizq.net. Retrieved 2010-08-22.
6. "Defining and Measuring Cloud Elasticity". KIT Software Quality Departement. Retrieved 13 August 2011.
7. "The Internet Cloud". Thestandard.com. Retrieved 2010-08-22.
8. Mohammad Hamdaqa, Tassos Livogiannis, Ladan Tahvildari: A Reference Model for Developing Cloud Applications. CLOSER 2011: 98-103
9. "Microsoft Plans 'Cloud' Operating System". Nytimes.com. Retrieved 2011-08-20.
10. "Cloud Net Directory. Retrieved 2010-03-01"